\documentclass[aps,nofootinbib,amsmath,prd,
twocolumn,
showpacs,superscriptaddress,groupedaddress]{revtex4}

\usepackage{graphicx}  
\usepackage{dcolumn}   
\usepackage{bm}        
\usepackage{amssymb}   

\usepackage{upgreek}

\begin{document}

\mbox{}

\title{Effective models of membranes from symmetry breaking}

\author{Omar Zanusso}
\email{O.Zanusso@science.ru.nl}
\affiliation{
Institute for Mathematics, Astrophysics and Particle Physics,
Radboud University Nijmegen,
Heyendaalseweg 135, 6525 AJ Nijmegen, The Netherlands}


\begin{abstract}
We show how to obtain all the models of the continuous
description of membranes by constructing the appropriate non-linear realizations
of the Euclidean symmetries of the embedding. The procedure has the advantage of giving a unified formalism with which the models are generated
and highlights the relevant order parameters in each phase.
We use our findings to investigate a fluid description of both tethered and hexatic membranes,
showing that both the melting and the loss of local order induce long range interactions in the high temperature fluid phase.
The results can be used to understand the appearance of intrinsic ripples in crystalline membranes in a thermal bath.
\end{abstract}

\maketitle

\section{Introduction.}

An effective model of membrane is the continuous description of an intrinsically two-dimensional object.
Examples of nature's realized membranes are ubiquitous and are of both biological and non biological nature.
On the biological side, cells' membranes,
which are characterized by both fluid (amphiphilic bilayers) and crystalline (cytoskeleton) properties, are a significant example \cite{Bowick:1999rk}.
On the other hand, the interest in non-biological two-dimensional crystals has been renewed in recent years
due to the discovery of graphene and similar materials with enormous potential technological applications \cite{CastroNeto:2009zz}.
From both a theoretical and a phenomenological point of view, it is of interest to investigate the phase-diagram of these structures
to gain better knowledge of their behavior under thermal and mechanical stresses.

In a statistical mechanics framework the phase-diagram can be obtained assigning a microscopic (bare) Hamiltonian to the membrane model
and computing the corresponding free-energy via a path-integral formalism.
The formalism allows to identify the critical points of the model, which separate the salient mechanical phases of the membrane.
For the sake of this introduction, it is useful to consider the membrane as an effective description of a layer of fundamental constituents (monomers)
linked together by a bonding interaction of fixed connectivity (crystal).
The phase-diagram of a theory of membranes is generally very complex because two-dimensional geometry allows for various
definitions of local order with corresponding order-parameters.

In a continuous formulation the membrane is the image of a map
\begin{eqnarray}\label{membrane}
 r:\mathbb{R}^d\to\mathbb{R}^D\,,
\end{eqnarray}
where in the physically interesting case of a two-dimensional membrane embedded in three-dimensional space one has $d=2$ and $D=3$,
but for the sake of generality in most of this work we will leave the couple $(d,D)$ general.
As a general requirement, we demand any membrane model to be invariant under the isometries of the embedding
\begin{eqnarray}\label{isometry}
 r^\mu \to R(\alpha)^\mu{}_\nu r^\nu + b^\mu\,,
\end{eqnarray}
where $R(\alpha)$ is any $D$-dimensional rotation parametrized by some angles $\alpha$ and $b$ is a general translation vector in $\mathbb{R}^D$.
The continuous formulation is of course only an effective description of the fundamental monomers' interactions.
If $a$ is the typical intra-monomer length, the effective continuous description will generally work well for
$h c \beta\ll a$ where $\beta=1/k_B T$ is the inverse of the temperature in units of the Boltzmann constant.
In fact, the examples cited above are characterized by $a\sim 1\,{\rm nm}$ (bilayers and graphene) and $a\sim 1\,\upmu {\rm m}$ (cytoskeleton),
and admit an accurate continuous description at the higher length scales of $10a-100a$.

We refer to the study of statistical field theories of \eqref{membrane} as the Kosterlitz-Thouless-Nelson-Halperin-Young (KTNHY) description \cite{Kosterlitz:1973xp,NH,Y}.
To summarize it, it is convenient to introduce the binding energy scale $E_b$ which might or might not be of order $1/a$.
One can distinguish low- $T\ll E_b/k_B$, high- $T\gg E_b/k_B$ and intermediate-temperature $T\simeq E_b/k_B$ phases.
Ideally, at low temperatures the bindings are intact and the monomers are in their ``natural'' crystalline phase.
In this phase the local connectivity is unaltered, thus the thermal fluctuations are elastic and can affect only the bindings' lengths.
At high temperatures the bindings melt and the membrane undergoes a fluid phase
that is characterized by the absence of both local connectivity and order.
In the intermediate temperatures phase, the binding interactions are assumed to be relevant, but not dominant.
The intermediate regime describes the melting and the local connectivity plays the role of the order parameter \cite{nelson,david}.

In this paper we want to show how to derive the three main effective models of the KTNHY description in a unified formalism
by considering the breaking of the global symmetries of the embedding space $\mathbb{R}^D$ due to the presence of the membrane \cite{West:2000hr}.
For this purpose, we will use the formalism of the Maurer-Cartan form (MCF) which proves very efficient in the investigation
of symmetry breaking patterns.
The MCF formalism has the advantage of clearly identifying the order parameters of the broken symmetries through a coset construction.
The order parameters obtained in this way thus enjoy a formal definition as outlined in \cite{Weinberg:1996kr}.

\section{The coset construction and ${\rm ISO}\!\left(D\right)$.}

In this section we want to briefly review the use of the MCF in the construction of a coset model
that represents the symmetry breaking pattern of a theory of membranes \eqref{membrane}
and involves the breaking of both internal and embedding symmetries \cite{gomis}.
The interested reader can find more details in the excellent recent review in \cite{Delacretaz:2014oxa}.
Let $G$ be the symmetry group of a membrane theory and let it be the direct product of ${\rm ISO}\!\left(D\right)$,
which is the Euclidean isometry group of the embedding \eqref{isometry},
and some compact internal symmetry group $G_{\rm int}$ related to the crystalline structure of the membrane
\begin{eqnarray}
 G = {\rm ISO}\!\left(D\right) \times G_{\rm int}\,.
\end{eqnarray}
The group $G$ is thus generated by the standard Euclidean translation $P_\mu$ and rotation $J_{\mu\nu}$ operators
that enjoy the algebra
\begin{equation}
\begin{tabular}{ c l l }
  $\left[J_{\mu\nu},J_{\rho\sigma}\right]$ & $=$ & $i\left(\delta_{\mu\rho}J_{\nu\sigma} + {\rm perm.}
  \right)$\,, \\
  $\left[J_{\mu\nu},P_{\rho}\right]$ & $=$ & $i\left(\delta_{\mu\rho}P_\nu-\delta_{\nu\rho}P_\mu\right)$\,, \\
  $\left[P_{\mu},P_{\rho}\right]$ & $=$ & $0$\,,
\end{tabular}
\end{equation}
as well as by some generator $T_m$ of the internal symmetry group $G_{\rm int}$ such that
\begin{equation}
\begin{tabular}{ l l l }
  $\left[T_m,J_{\mu\nu}\right]=0$\,, & \qquad & $\left[T_m,P_\mu\right]=0$\,.
\end{tabular}
\end{equation}

We assume that a general membrane configuration breaks $G$ spontaneously to the unbroken subgroup $H \subset G$.
We also assume that $H$ might be non-compact,
implying that some translations of ${\rm ISO}\!\left(D\right)$ might not be broken by the physical configurations.
Excluding the unbroken translations from $H$, it is possible to identify a compact subgroup $H_0 \subset H$, which is generated by the unbroken
rotations and internal symmetries of $G$.
In this case an effective membrane model can be obtained as the coset
\begin{eqnarray}\label{general_coset}
 G\big \slash H_0\,.
\end{eqnarray}
From the general theory of cosets, the effective model will be manifestly invariant under the subgroup $H_0$,
but it will maintain the full $G$ symmetry, albeit it will be realized, at least partly, non-linearly.

Any element of the coset \eqref{general_coset} can be parametrized by an equivalence class of elements of $G$ under the right-action of $H_0$.
It is convenient to choose a representative for the equivalence class of the form
\begin{eqnarray}\label{omega}
 \omega(r,\xi) = {\rm e}^{i r^\mu P_\mu} \, {\rm e}^{i \xi^A X_A}\,,
\end{eqnarray}
where $X_A$ are all the broken generators of both embedding and internal symmetries.
The equivalence class is thus related to $\omega(r,\xi)$ as
\begin{eqnarray}
 \left[\omega(r,\xi)\right] = \left\{\omega(r,\xi)\cdot h;\,\forall h\in H_0\right\}\,,
\end{eqnarray}
and by construction any of its elements has the same physical field content.
In general, $\omega(r,\xi)$ is parametrized by the field content of the model $r^\mu$
and by the Goldstone fields $\xi^A$ associated with the broken symmetries.
By construction, the element \eqref{omega} transforms under $G$ by the left-action of a generic element $g\in G$ as
$ \omega(r,\xi)\to g\cdot\omega(r,\xi)$.
The action of $G$ may bring one coset representative into another inequivalent one,
thus defining implicitly the transformation properties of $r^\mu$ and $\xi^A$ via
\begin{eqnarray}
 \omega(r,\xi) \to g\cdot\omega(r,\xi) \equiv \omega(r',\xi')\cdot h' \,,
\end{eqnarray}
where $h'$ belongs to the unbroken $H_0$ and returns the representative to the form \eqref{omega}.

The effective model for the field arguments of \eqref{omega} can be constructed using the MCF,
which is defined as
\begin{eqnarray}
 L=\omega(r,\xi)^{-1}{\rm d} \omega(r,\xi)  \,,
\end{eqnarray}
which by construction belongs to the algebra of $G$.
Once the physical content of $L$ is identified via its algebra components an effective Hamiltonian for
the theory with broken symmetries can be written down.
In the following we will follow this method, adopting physically motivated symmetry breaking patterns
for all the models of the KTNHY description.

\section{The tethered membrane.}

\begin{figure}[ht]
 \includegraphics[width=4cm]{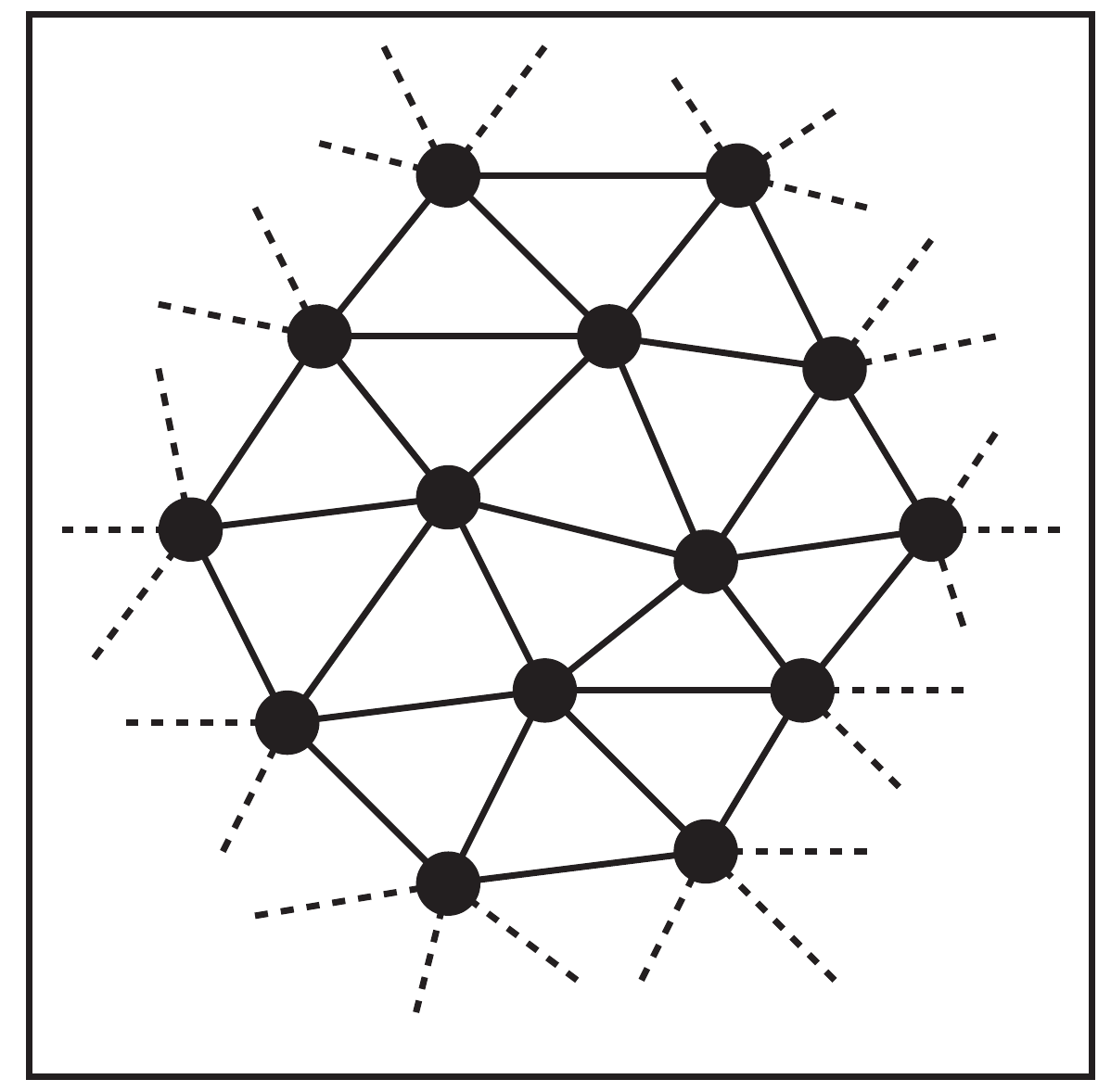}
 \caption{
 Tethered membrane.
 Each monomer (black dot) is regarded as a point-like object that breaks fully the translational invariance of the embedding.
 In the crystalline phase the bindings (lines) are intact, it is relevant to deform the membrane by displacing each single monomer in any direction.
 }
 \label{tethered_fig}
\end{figure}

At low temperatures the bonding interactions among the monomers are intact and favor a crystalline structure \cite{nelson,tethered}.
In this situation each monomer fluctuates under mechanical stresses as a point-like object as in Fig.\ \ref{tethered_fig}.
A point-like object has a definite position in space, therefore its configurations break all translations, while leaving the Euclidean rotations invariant.
For simplicity, in this Section we neglect any internal symmetry group of the monomers' lattice.
The full group of symmetries is thus $G={\rm ISO}\!\left(D\right)$, while the unbroken symmetries are $H_0={\rm SO}\!\left(D\right)$.
The coset is thus
\begin{eqnarray}
 {\rm ISO}\!\left(D\right) \big \slash {\rm SO}\!\left(D\right)\,.
\end{eqnarray}
As representative of the coset we choose the simple
\begin{eqnarray}
 \omega = {\rm e}^{i r^\mu P_\mu}\,.
\end{eqnarray}
We now introduce a set of coordinates $x^\alpha$ of $\mathbb{R}^d$ and compute the MCF as
\begin{eqnarray}
 L_\alpha = \omega^{-1} \partial_\alpha \omega  = i \partial_\alpha r^\mu P_\mu\,.
\end{eqnarray}
The components $e^\mu_\alpha\equiv\partial_\alpha r^\mu$ of the MCF are the Goldstone fields of the broken translations
and are formally defined as the order parameters of the model in the sense of \cite{Weinberg:1996kr}.
By construction, the components $e^\mu_\alpha$ transform linearly under the ${\rm SO}\!\left(D\right)$ rotation subgroup of \eqref{isometry},
but are scalars under the broken translations:
\begin{equation}\label{transformations_tethered}
 \begin{split}
  \begin{tabular}{ l l l }
  $e^\mu_\alpha \to R(\alpha)^\mu{}_\nu\, e^\nu_\alpha$ & \,for\, & $r^\mu \to R(\alpha)^\mu{}_\nu\, r^\nu$\,,\\
  $e^\mu_\alpha \to e^\mu_\alpha$ & \,for\, & $r^\mu \to r^\mu + b^\mu$\,.
\end{tabular}
 \end{split}
\end{equation}

A general Hamiltonian for a crystalline phase can thus be constructed only from $e^\mu_\alpha$ and its derivatives.
Up to fourth order in a derivative expansion and expressing $e^\mu_\alpha$ in terms of the membrane configuration $r^\mu$,
the most general Hamiltonian is
\begin{equation}\label{tethered}
\begin{split}
 \beta {\cal H}_t\left[r\right] =&
 \int\!{\rm d}^dx\,\Bigl\{
 \epsilon_0
 +\frac{\mu_t}{2} (\partial_\alpha r^\mu)^2
 +\frac{\kappa_t}{2} (\partial^2 r^\mu)^2
 \\
 &\quad
 +u(\partial_\alpha r^\mu \partial_\beta r^\mu)^2
 +v(\partial_\alpha r^\mu \partial_\alpha r^\mu)^2
 \Bigr\}\,,
\end{split}
\end{equation}
which is known as the tethered membrane model \cite{nelson,tethered}.
We neglected possible boundary terms and adopted a rather general notation for the couplings that have been introduced:
$\epsilon_0$ is a chemical potential for the monomer number that can be neglected at fixed membrane volume,
$\mu_t$ is the surface tensions, $\kappa_t$ is the extrinsic rigidity,
and $u$ and $v$ are Lam\'e coefficients parametrizing the elastic properties of the membrane.

The fields $e^\mu_\alpha$ are the order parameters of the model and distinguish the possible mechanical phases of the theory.
To be more precise we introduce a path integral with which Boltzmann averages of any operator ${\cal O}[r]$ can be computed as
\begin{eqnarray}
 \left<{\cal O}[r]\right> = 
 \int {\rm D}r\,{\cal O}[r]\,{\rm e}^{-\beta {\cal H}_t\left[r\right]}
 \,,
\end{eqnarray}
where we defined a properly normalized measure ${\rm D}r$ that is invariant under \eqref{transformations_tethered}.
Two mechanical phases can be roughly distinguished as
\begin{equation}\label{phases_tethered}
 \begin{split}
  \begin{tabular}{ l l l }
  $\left<\int\!{\rm d}^dx\,e_\alpha^2\right> \big \slash \int\!{\rm d}^dx =0$ & \qquad & crumpled\,,\\
  $\left<\int\!{\rm d}^dx\,e_\alpha^2\right> \big \slash \int\!{\rm d}^dx \neq 0$ & \qquad & flat\,,
\end{tabular}
 \end{split}
\end{equation}
where $e_\alpha^2$ indicates the square of the order parameter $e^2=\sum_\mu e^\mu_\alpha e^\mu_\alpha$
(no summation over $\alpha$).
Each order parameter $e_\alpha^\mu$ enjoys a separate treatment, thus 
anisotropic phases in which the membrane is crumpled along some directions
while extended along the others are possible \cite{tubule}.
For a more detailed description of these phases we refer to \cite{Bowick:1999rk}.
The order of the phase-transition separating crumpled and flat phases is still subject to investigation
and is potentially very relevant for applications in the developing physics of graphene.
The most recent non-perturbative treatment as well as lattice simulations suggests that the crumpled-to-flat transition
is of first order \cite{Essafi:2014dla,troster}.

\section{The fluid membrane.}\label{f-section}

\begin{figure}[ht]
 \includegraphics[width=4cm]{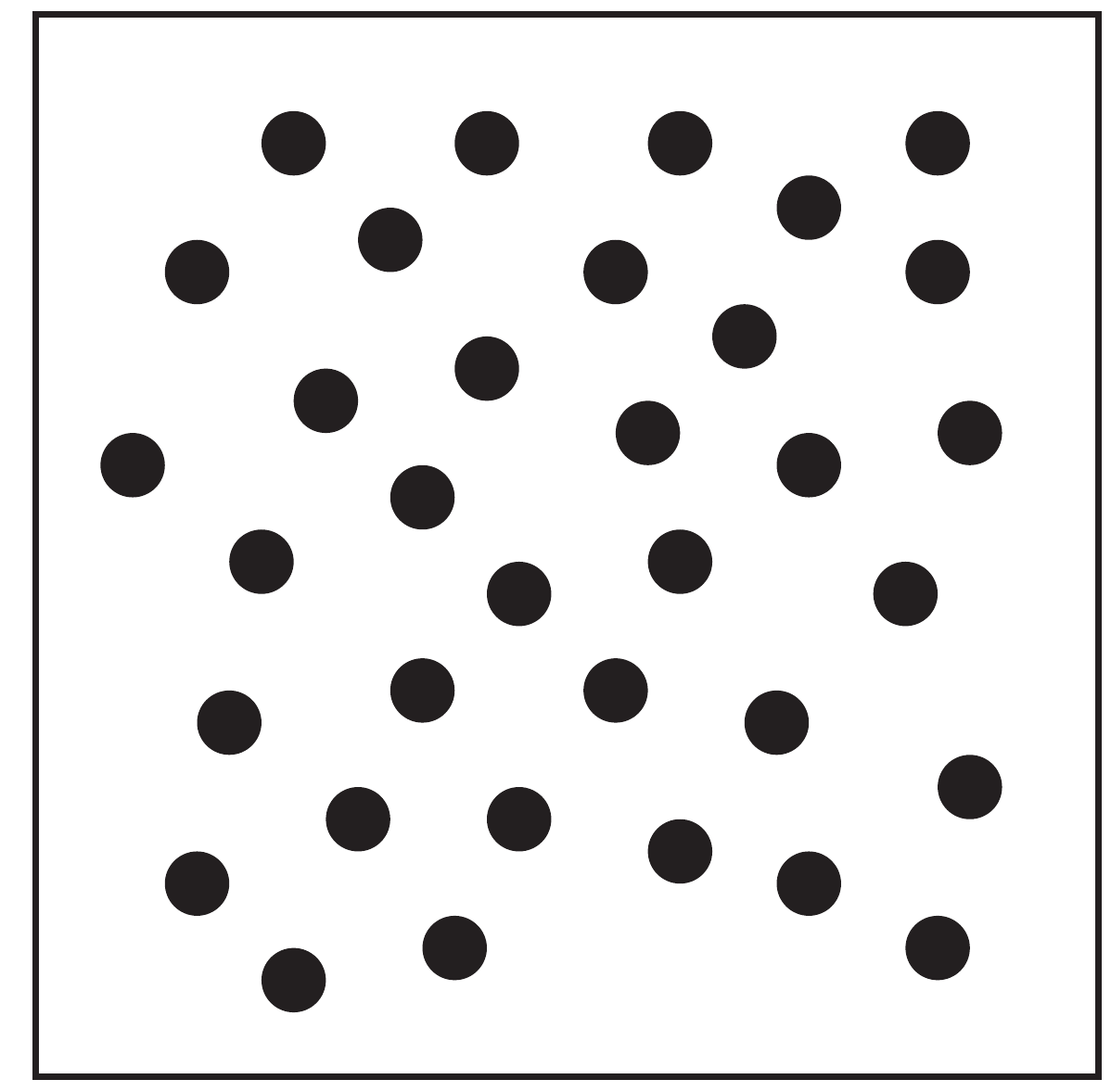}
 \caption{
 Fluid membrane.
 The monomers are free to diffuse, so the in-plane displacements do not change the (macroscopic) continuous description of the membrane.
 }
 \label{fluid_fig}
\end{figure}

At high temperatures the bindings among the monomers melt, which are then free to diffuse along the membrane \cite{fluid}.
At macroscopic equilibrium, two membrane configurations cannot be distinguished if a monomer is translated along the tangential directions of the membrane as in Fig.\ \ref{fluid_fig}.
To picture the phase, it is convenient to consider the infinitesimal plaquette of the membrane in position $r^\mu$ defined by
$r^\mu+{\rm d}r^\mu=r^\mu+\partial_\alpha r^\mu {\rm d}x^\alpha$.
Infinitesimally, the plaquette breaks spontaneously the translations perpendicular to the plane generated by $\partial_\alpha r^\mu {\rm d}x^\alpha$
and the rotations that do not leave the same plane invariant.
Again, we shall neglect any possible internal symmetry group in the course of this Section.

The (compact) unbroken subgroup for the fluid membrane is therefore $H_0={\rm SO}\!\left(d\right)\times {\rm SO}\!\left(D-d\right)$
and includes both the ${\rm SO}\!\left(d\right)$ subgroup of local frame rotations on the plaquette
and the subgroup ${\rm SO}\!\left(D-d\right)$ of rotations of its normal space.
The coset is thus
\begin{eqnarray}
 {\rm ISO}\!\left(D\right) \big \slash {\rm SO}\!\left(d\right)\times {\rm SO}\!\left(D-d\right)\,.
\end{eqnarray}
In this Section we will follow closely the construction of \cite{Delacretaz:2014oxa},
which was developed in a different context, but applies here with minor modifications.
Before choosing a coset representative, it is convenient to switch to a system of coordinates in $\mathbb{R}^D$
that locally aligns the first $d$ axes with the tangents of the membranes
and the remaining $D-d$ ones with their normal space.
In the new system of coordinates the membrane is described by the couple $r^\mu=(r^\alpha,r^i)$.
The coset representative is then chosen to include all the broken generators as well as the unbroken translations
\begin{eqnarray}\label{representative_fluid}
 \omega={\rm e}^{i r^\alpha P_\alpha+i r^i P_i}{\rm e}^{i\xi^{\alpha i}J_{\alpha i}}\,.
\end{eqnarray}
For later use, we introduce a further set of coordinates $y^a$ on the membrane whose purpose will become clear below.

In order to correctly identify the field content of the model,
we first choose a general parametrization of the MCF of the form
\begin{eqnarray}\label{form_fluid}
 L_a = i e_a{}^\alpha ( P_\alpha + \nabla_\alpha \pi^i P_i + \nabla_\alpha \xi^{\gamma i}J_{\gamma i} + A_\alpha{}^{\beta\gamma}J_{\beta\gamma})\,,
\end{eqnarray}
where both $\pi^i$ and $\xi^{\gamma i}$ have to be thought of as Goldstone fields of the broken translations and rotations, respectively,
while $A_\alpha{}^{\beta\gamma}$ are unimportant tensors corresponding to unbroken local rotations on the membrane.
The components $\nabla_\alpha \pi^i$ correspond to the broken normal translations.
Whenever translations are broken in a system the usual counting of the degrees of freedom of the Goldstone theorem
is more subtle, as there is not necessarily one Goldstone field for each broken symmetry generator.
The mismatch in the counting follows what is known as the inverse Higgs mechanism.
The redundant Goldstone fields can be eliminated by imposing a so-called inverse Higgs constraint \cite{Ivanov:1975zq}.
The simplest possible constraint for our system is
\begin{equation}\label{IHC}
 \begin{split}
 \nabla_\alpha \pi^i=0\,,
 \end{split}
\end{equation}
while all other possible choices can be related to this one by a redefinition of the couplings.
The solution of \eqref{IHC} can then be used to express the Goldstone fields of the broken translations $\pi^i$
in terms of those of the broken rotations $\xi^{\alpha i}$.
Through this mechanism the correct counting of the Goldstone fields is achieved.
From a physical point of view, it is possible to interpret the theory constrained by \eqref{IHC} in various complementary ways \cite{Delacretaz:2014oxa}.
In this context, we can understand the constrained theory as the one in which the modes $\pi^i$
have been effectively integrated out and refer to Sect.\ \ref{t-f-section} for further insights on this interpretation.

Using \eqref{representative_fluid} the explicit computation of the components of the form \eqref{form_fluid} corresponding to the translations gives
\begin{equation}\label{form_fluid_translations}
 \begin{split}
  e_a{}^\alpha &= \partial_a r^\mu R(\xi)_\mu{}^\alpha\\
  e_a{}^\alpha \nabla_\alpha \pi^i &= \partial_a r^\mu R(\xi)_\mu{}^i\,,
 \end{split}
\end{equation}
where we introduced the orthogonal rotation $R(\xi)$ parametrized by the Goldstone fields of the broken rotations.
The condition \eqref{IHC} simply implies
\begin{eqnarray}\label{inverse_higgs_fluid}
 \partial_a r^\mu R(\xi)_\mu{}^i=0\,,
\end{eqnarray}
which defines the orthonormal basis $R(\xi)_\mu{}^i = n^i_\mu$.
The normal basis is obviously directly related to the Goldstone fields of the broken rotations.

The fields $e_a{}^\alpha$ transform covariantly under the unbroken rotations and can be used as frames on the membrane.
To show this and give a geometrical meaning to the construction we introduce the metric
\begin{eqnarray}\label{metric}
 g_{ab} = e_a{}^\alpha e_b{}^\beta \delta_{\alpha\beta}\,.
\end{eqnarray}
Using \eqref{form_fluid_translations} and the invariance of the embedding space metric $\delta_{\alpha\beta}$ under orthogonal rotations
it is very easy to see that the metric corresponds to the induced metric on the membrane $g_{ab}=\partial_a r^\mu \partial_b r^\nu \delta_{\mu\nu}$
using the new coordinate patch $y^a$.
From now on, Latin indices from the beginning of the alphabet will be raised and lowered using $g_{ab}$ and its inverse.
The derivatives of the Goldstone fields can be interpreted geometrically too.
We first compute the components of the form \eqref{form_fluid} corresponding to the broken rotations
\begin{eqnarray}\label{form_fluid_rotations}
 e_a{}^\alpha\nabla_\alpha \xi^{\gamma i}
 &=&  [R(\xi)^{-1}\partial_a R(\xi)]^{\gamma i}\,.
\end{eqnarray}
Introducing the inverse frame $e{}_\alpha{}^a\equiv e^{-1}{}_\alpha{}^a$, it can be directly inverted and related to the normal basis as
\begin{eqnarray}\label{eqn1}
 \nabla_\alpha \xi^{\gamma i}
 &=&
 e_\alpha{}^a R(\xi)^{-1}{}^{\gamma \mu} \partial_a n^i_\mu\,.
\end{eqnarray}
This expression should be compared with the definition of extrinsic curvature of a membrane,
which we manipulate using the constraint \eqref{inverse_higgs_fluid} as
\begin{eqnarray}\label{eqn2}
 K^i_{ab} \equiv  \partial_b r^\mu \partial_a n^i_\mu = e_b{}^\beta R(\xi)_\beta{}^\mu{} \partial_a n^i_\mu\,.
\end{eqnarray}
Combining \eqref{eqn1} and \eqref{eqn2} we immediately see
that the derivatives of the Goldstone fields of the broken rotations are directly related to the extrinsic curvatures as
\begin{eqnarray}
 K^i_{ab}
 &=& e_a{}^\alpha e_b{}^\beta  \nabla_\alpha \xi^i{}_\beta\,.
\end{eqnarray}

The most general Hamiltonian that can be constructed from the MCF can only be a function of
$e_a{}^\alpha$, $\nabla_\alpha \xi^i{}_\beta$ and their derivatives.
Due to the geometrical meaning of these quantities it can thus be written as a function of metric $g_{ab}$ and extrinsic curvatures $K^i_{ab}$.
The Hamiltonian will also be reparametrization invariant for transformations of the arbitrary coordinate $y^a$.
Neglecting boundary terms, the most general invariant Hamiltonian up to second order in the derivatives is
\begin{eqnarray}\label{fluid}
 \beta {\cal H}_f\left[r\right] &=&
 \int\!{\rm d}^dy\,\sqrt{g}\,\Bigl\{
 \mu_f
 +\frac{\kappa_f}{2} K^2
 +\frac{\bar{\kappa}}{2} R
 \Bigr\}\,,
\end{eqnarray}
where $K^2$ is the square $K^2=K^i K^i$ of the traces of the extrinsic curvatures $K^i=g^{ab} K^i_{ab}$,
$g^{ab}$ is the inverse of the induced metric \eqref{metric},
$R=g^{ab}R_{ab}=g^{ab}R_{ca}{}^c{}_b$ is the (intrinsic) curvature scalar,
$R_{ab}{}^c{}_d$ is the Riemann tensor defined by $\left[\nabla_a,\nabla_b\right]v^c=R_{ab}{}^c{}_dv^d$,
$\nabla_a$ is the metric-compatible connection defined by $\nabla_a v^b=\partial_a v^b+\Gamma_a{}^b{}_c v^c$
with the Christoffel symbols obtained from the induced metric
$\Gamma_a{}^b{}_c =\frac{1}{2}g^{bd}(\partial_a g_{dc} +\partial_c g_{ad} - \partial_d g_{ac})$
and $g=\det{g_{ab}}$ is the determinant of the metric.
This is the so-called fluid membrane model, which is also well known in string theory \cite{polyakov}.
The coupling $\mu_f$ is the surface tensions, $\kappa_f$ is the (fluid) extrinsic rigidity,
and $\bar{\kappa}$ is known as the Gaussian rigidity of the membrane.

In the fluid model reparametrization invariance has a clear physical origin that we shall outline before concluding the Section.
The general mechanical deformation of the fluid membrane would perturb it along
both tangential and normal directions like in the tethered model of the previous Section.
It can be parametrized as
\begin{equation}
 r^\mu \to r^\mu + \nu^a \partial_a r^\mu + \nu^i n^\mu_i\,.
\end{equation}
From the discussion above, however, it is clear that at its equilibrium the fluid membrane macroscopic state is unaffected
by tangential deformations $\nu^a$ corresponding to translations of the monomers along the membrane.
Reparametrization invariance ensures that the transformations $\nu^a$ can be absorbed by a corresponding
reparametrization of the coordinates $y^a$ which is infinitesimally parametrized by a vector field tangent to the membrane itself.
A path-integral can be constructed using \eqref{fluid} as
\begin{eqnarray}\label{pi_fluid}
 \left<{\cal O}[r]\right> = 
 \int {\rm D}r_{\rm rep}\,{\cal O}[r]\,{\rm e}^{-\beta{\cal H}_f\left[r\right]-\beta{\cal H}_{\rm g.f.}}
 \,,
\end{eqnarray}
where ${\rm D}r_{\rm rep}$ is a new normalized measure that respects reparametrization invariance
and ${\cal H}_{\rm g.f.}$ is an opportune gauge fixing term.
When constructing \eqref{pi_fluid} it is often convenient to adopt the background field method
and choose the physical gauge for which $\nu^a=0$, which is analogous to the Landau gauge of Yang-Mills theories.
Albeit quite different from the point of view of the coset construction,
the fluid membrane shares the same mechanical phases of the tethered membrane
and the same discussion of \eqref{phases_tethered} applies \cite{Codello:2011yf}.
It can however manifest a phase-transition of different order.
We will return to this point in the next Section, highlighting a substantial difference between the two models.

\section{From tethered to fluid membrane.}\label{t-f-section}

In this Section we attempt a fluid model description of the tethered membrane.
We will achieve it by properly integrating the appropriate Goldstone fields.
For simplicity we will deal with the Hamiltonian of the tethered model \eqref{tethered}
truncated to the second order of the derivative expansion
\begin{eqnarray}
 \beta{\cal H}_t\left[\mathbf{r}\right] &=&
 \int\!{\rm d}^dx\,\Bigl\{
 \epsilon_0
 +\frac{\mu_t}{2} (\partial_\alpha r^\mu)^2\Bigr\}\,,
\end{eqnarray}
which has to be regarded as a toy model that neglects both the rigidity and the elastic properties.
We notice that this Hamiltonian is evidently not reparametrization invariant
and this is the main difference from the fluid model of the previous Section.
In this sense we interpret the coordinates $x^\alpha$ as a fiducial set that describes the position
of the monomers on the membrane and has a special role
if compared to the arbitrary coordinate patch $y^a$ introduced for the fluid model.

The task is now to rewrite this action in terms of the fields appearing in the nonlinear realization \eqref{form_fluid}
\emph{before} imposing the inverse Higgs constraint.
This can be done performing the symmetry transformation
\begin{equation}
\partial_a r^\mu \to R(\xi)^\mu{}_\nu \partial_a r^\nu\,, 
\end{equation}
however it is also necessary to perform a coordinate change from the ``crystalline'' patch $x^\alpha$
to the arbitrary coordinate patch $y^a$
with the transformation
\begin{equation}
 \partial x^\alpha  \slash \partial y^a = e_a{}^\alpha\,.
\end{equation}
The result of the manipulations is
\begin{eqnarray}\label{tethered_small}
 \beta{\cal H}_t\left[r\right] &=&
 \int\!{\rm d}^dy\sqrt{g}\,\Bigl\{
 \epsilon_0+\frac{d}{2}\mu_t
 +\frac{\mu_t}{2} (\nabla_a\pi^i)^2\Bigr\}\,,
\end{eqnarray}
where $g_{ab}$ is again the induced metric \eqref{metric} and $\pi^i$ was introduced in \eqref{form_fluid}.

The task is now to integrate the field $\pi^i$, which here plays the role of the order parameter
for the breaking of the symmetry
$${\rm SO}\!\left(D\right)\to {\rm SO}\!\left(d\right)\times {\rm SO}\!\left(D-d\right)\,,$$
which characterizes the compact symmetry content of the transition from the tethered to the fluid descriptions.
The fields $\pi^i$ transform as $D-d$ scalars under ${\rm SO}\!\left(d\right)$.
The path integration of the Goldstone fields $\pi^i$ is performed using the properly normalized measure ${\rm D}\pi$ as
\begin{eqnarray}\label{pi}
 {\rm e}^{-\beta{\cal H}_{\rm t-f}\left[r\right]} = 
 \int {\rm D}\pi\,{\rm e}^{-\beta{\cal H}_t\left[r,\pi\right]}
 \,,
\end{eqnarray}
and provides a physical realization of the inverse Higgs mechanism \eqref{IHC} of Sect.\ \ref{f-section}.
For simplicity we shall consider the physical model with $d=2$ and $D=3$.
In this case $\pi^i=\pi$ is a single scalar and
thus an integration of its kinetic term in \eqref{tethered_small}
that preserves reparametrization invariance gives the well known Liouville action
(see for example \cite{Codello:2010mj}).
We obtain an effective fluid Hamiltonian for the tethered model as
\begin{eqnarray}\label{t-f}
 \beta{\cal H}_{\rm t-f}\left[r\right] &=&
 \int\!{\rm d}^2y\sqrt{g}\,\Bigl\{
 \mu_R
 - \frac{1}{96\pi} R\frac{1}{\Delta}R
 \Bigr\}\,,
\end{eqnarray}
where we introduced a renormalized tension
$$\mu_R=\epsilon_0+\frac{d}{2}\mu_t$$
that could be easily evinced from \eqref{tethered_small},
the Laplacian operator $\Delta=-\nabla^a\partial_a$,
and the curvature scalar of the induced metric $R$.
The new effective Hamiltonian enjoys reparametrization invariance as required for a fluid description.
Interestingly, the integration of the local order parameter controlling the breaking of the symmetry interfacing a tethered-fluid transition
gives rise to long range interactions among the intrinsic curvatures of the fluid phase.

Long range interactions such as that of \eqref{t-f} are of particular importance in two-dimensional systems,
as they are known to provide a way out of the Mermin-Wagner theorem \cite{MW} and are responsible for the non-triviality
of systems that undergo a Kosterlitz-Thouless type of transitions \cite{Kosterlitz:1973xp}.
The sole presence of the long range interactions casts the question on whether the tethered and fluid models
should share the same phase-diagram.
A complete answer to this question is still unknown, even though in the past it has been investigated at length using lattice methods \cite{lattice,troster}.
To gain a better insight physical quantities should be compared among the different phases of the two models at least qualitatively.
The advantage of the formalism described in this paper is that it makes possible a quantitative comparison
of the results for the phase-diagrams of the two models.

\begin{figure}[tph]
 \includegraphics[width=4.5cm]{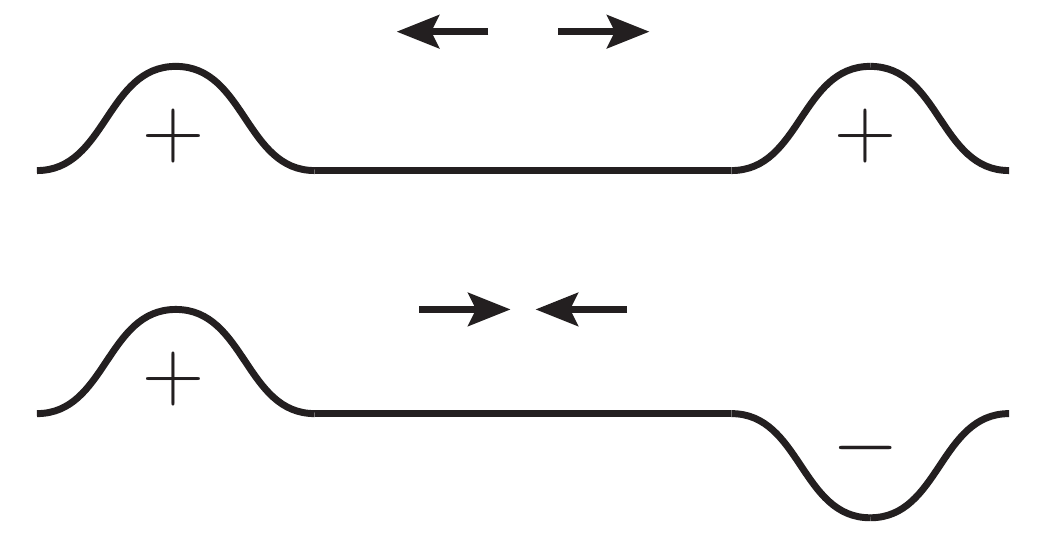}
 \caption{
 Naive visualization of the long range interactions in the fluid description of tethered and hexatic membranes.
 Configurations of the membrane with alternating sign of the scalar curvature are energetically favored.
 }
 \label{longrange}
\end{figure}

One particularly important observable in the continuum is the fractal (Hausdorff) dimension $d_f$ of the membrane \cite{david},
which here represents how fuzzy a crystalline membrane becomes in the process of melting.
Given \eqref{t-f} we can compute an estimate of the membrane's fractal dimension
using results originally developed for the hexatic model, which will be discussed in the next Section.
Using either the one-loop result of David et al.\ \cite{david} or the nonperturbative renormalization group result of \cite{Codello:2013pya}
we obtain
\begin{equation}\label{t-f-fractal}
 \begin{split}
  \begin{tabular}{ l l l }
  $d_f = 14$ & \qquad & one-loop\,,\\
  $d_f = 2.57$ & \qquad & non-perturbative\,.
\end{tabular}
 \end{split}
\end{equation}
The high numerical difference between the two numbers is due to the fact
that an expansion in the inverse coupling of the Liouville action is performed in the one-loop result,
making the non-perturbative result more reliable.
The physical understanding of these numbers goes as follows:
The Liouville interaction of \eqref{t-f} tends to attract intrinsic curvatures of different sign,
making the surface more and more fuzzy as different curvatures are packed together as hinted by Fig.\ \ref{longrange}.

The non-perturbative result resums the curvature interactions to a higher extent
and thus includes some higher-order screening that considerably lowers the value of $d_f$.
Since we are neglecting any internal symmetry the melting tethered surface under consideration is not characterized by local order.
Our result thus predicts that an unstructured crystal
undergoes a crinkled phase characterized by a definite fractal dimension
in the process of melting.
In the crinkled phase the membrane is characterized by intrinsic ripples of the surface.
The physics of ripples is relevant for the understanding of the stability of two-dimensional crystals \cite{fasolino}.

\section{The hexatic membrane.}

\begin{figure}
 \includegraphics[width=4cm]{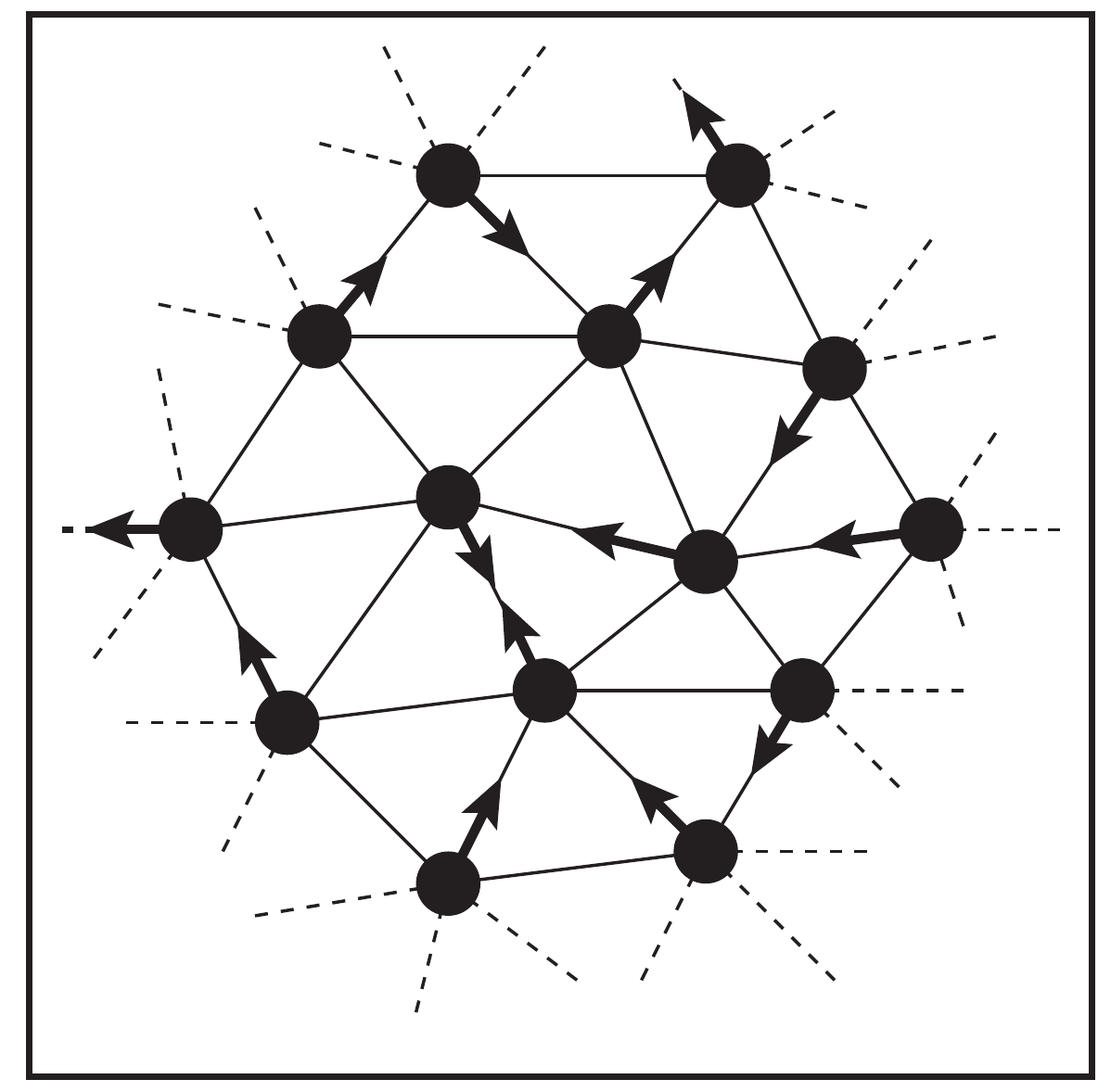}
 \caption{
 Hexatic membrane.
 The crystalline structure breaks the local rotations of the membrane.
 The arrows represent the order parameter for the associated phase-transition.
 A disordered phase is depicted.
 }
 \label{hexatic_fig}
\end{figure}

The hexatic membrane can be understood as a melting tethered membrane or as a fluid membrane
with the residual effect of the local order that breaks the internal rotations,
thus describing the intermediate temperature phase separating the crystalline low-temperature phase with the high-temperature fluid one \cite{david}.
In the physical case, the crystalline structure has to break the local ${\rm SO}\!\left(2\right)$ rotations
tangent to the membrane as depicted in Fig.\ \ref{hexatic_fig}.
To generalize this situation while maintaining the desired $d=2$ limit,
we assume that the crystal has an internal $G_{\rm int}={\rm SO}\!\left(d\right)_{\rm cr}$ symmetry with generators $T_A=S_{\alpha\beta}$
emerging from the continuous description of some discrete lattice group,
but more general breaking patterns can be studied in similar ways.
Differently from the previous Sections, the full group of symmetries is enhanced to
\begin{eqnarray}
 G = {\rm ISO}\!\left(D\right)\times {\rm SO}\!\left(d\right)_{\rm cr}\,.
\end{eqnarray}
The local rotations on the membrane with generators $J_{\alpha\beta}$ are broken,
but a combined rotation of the membrane and lattice leaves the configuration invariant.
We thus require that the subgroup
\begin{eqnarray}
 {\rm SO}\!\left(d\right)\times {\rm SO}\!\left(d\right)_{\rm cr} \subset {\rm ISO}\!\left(D\right)\times {\rm SO}\!\left(d\right)_{\rm cr}
\end{eqnarray}
is broken to the diagonal ${\rm SO}\!\left(d\right)$ generated by $M_{\alpha\beta}=J_{\alpha\beta}-S_{\alpha\beta}$.

As coset representative we choose a structure that enhances the breaking pattern \eqref{form_fluid} to the internal rotations
\begin{eqnarray}
 \omega ={\rm e}^{i r^\alpha P_\alpha+i r^i P_i}{\rm e}^{i\xi^{\alpha i}J_{\alpha i}+(i/2)\xi^{\alpha\beta}J_{\alpha\beta}}\,.
\end{eqnarray}
The MCF is then parametrized as
\begin{equation}\nonumber
 \begin{split}
 L_a = &\, i e_a{}^\alpha \Bigl( P_\alpha + \nabla_\alpha \pi^i P_i
 + \nabla_\alpha \xi^{\gamma i}J_{\gamma i} + \frac{1}{2}\nabla_\alpha\xi{}^{\beta\gamma}J_{\beta\gamma}\Bigr)\,,
 \end{split}
\end{equation}
where the fields $e_a{}^\alpha$ still play the role of frames for the membrane
and new Goldstone fields $\xi{}^{\beta\gamma}$ for the breaking of ${\rm SO}\!\left(d\right)$ appear.
The first three components are in a form equal to \eqref{form_fluid_translations} and \eqref{form_fluid_rotations},
while the remaining one can be computed as
\begin{equation}
 \begin{split}
 e_a{}^\alpha \nabla_\alpha \xi_\beta{}^\gamma &=  (R(\xi)^{-1}\partial_a R(\xi))_\beta{}^\gamma\,,
 \end{split}
\end{equation}
for a rotation $R(\xi)$ that is parametrized by $\xi_\beta{}^i$ and $\xi_\beta{}^\gamma$.

As in the fluid case, the constraint $ \nabla_\alpha \pi^i=0$ identifies a set of orthonormal vectors defined by $n^i_\mu= R(\xi)_\mu{}^i$.
In this case, however, an additional set of orthonormal tangent vector fields is identified
as $N^\gamma_\mu=R(\xi)_\mu{}^\gamma$ and describes the breaking of the local rotations.
They are related to the new Goldstone fields through the MCF as
\begin{eqnarray}
 \nabla_\alpha \xi^{\beta\gamma} &=&  e_\alpha{}^a R(\xi)^{-1}{}^{\beta\mu} \partial_a N^\gamma_\mu\,.
\end{eqnarray}
The construction of the Hamiltonian follows closely the fluid case, but includes the new order parameter for the broken rotations.
At second order in the derivative expansion the Hamiltonian contains \eqref{fluid} and new terms involving the Goldstone fields $\xi^{\alpha\beta}$
that can only be of the form
\begin{eqnarray}
 \beta{\cal H}\left[r,\xi\right] &=& \frac{K_A}{2} \int {\rm d}^dy \sqrt{g}\, I_{\alpha\beta\gamma\delta}\nabla_a\xi^{\alpha\beta} \nabla^a \xi^{\gamma\delta}\,,
\end{eqnarray}
where $I_{\alpha\beta\gamma\delta}$ is a tensor containing the details of the breaking of the combination of rotations and internal symmetries.
The coupling $K_A$ has been introduced to parametrize the strength of the new Goldstone fields interaction
and is known as hexatic rigidity.
For the case of diagonal breaking the symmetry requirement implies that we simply have
$I_{\alpha\beta\gamma\delta}=\delta_{\alpha\gamma}\delta_{\beta\delta}$.

In the physically interesting case of $d=2$ and $D=3$ there is only one local rotation on the membrane,
which is fully broken by almost all possible discrete lattice rotations.
The corresponding Goldstone field is an angular ${\rm SO}(2)$ variable $\theta=\xi^{12}$,
which we can associate with a tangent vector on the membrane defining $N^\alpha = \cos\theta \,e_1{}^\alpha+\sin\theta \,e_2{}^\alpha$.
When expressed in terms of $N^\alpha$, the Hamiltonian becomes a well-known formulation of the hexatic term
\begin{eqnarray}
 \beta{\cal H}_{\rm hex}\left[r,N\right] &=& \frac{K_A}{2} \int {\rm d}^dy \sqrt{g}\, \nabla_a N^\alpha \nabla^a N^\alpha\,.
\end{eqnarray}
At second order the vector $N$ appears quadratically and can thus be integrated away following the procedure of \cite{david}
that brings the integration over the new field in a form similar to \eqref{pi}.
The result of the integration is again a Liouville action contribution to the Hamiltonian of fluid model of the form
\begin{eqnarray}\label{hexatic}
 \beta{\cal H}_{\rm h-f}\left[\mathbf{r}\right] &=&
 -\frac{\bar{K}_A}{8} \int\!{\rm d}^2y\sqrt{g}\,
 R\frac{1}{\Delta}R
 \,,
\end{eqnarray}
where the coupling $\bar{K}_A$ underwent only a finite renormalization $\bar{K}_A=K_A-1/12\pi$.
The same comments of the previous section on the long range interactions induced by a Liouville action apply in this context as well.
The term \eqref{hexatic} is fundamental to circumvent the fact that a fluid membrane has an extended phase only at zero coupling
($1/\kappa_f=0$) \cite{PL}. The hexatic model, in fact, is known to display a non-trivial extended phase for a finite value of $\kappa_f$.
The fractal dimension of this phase is known perturbatively \cite{david} in an expansion in $1/\bar{K}_A$ as
\begin{equation}\label{sdim}
 d_f = 2+\frac{D(D-2)}{3\pi}\bar{K}^{-1}_A+{\cal O}\!\left(\bar{K}^{-2}_A\right)\,,
\end{equation}
and has been investigated non-perturbatively in \cite{Codello:2013pya}.

The nontrivial spectral dimension \eqref{sdim} is a manifestation of the fact that the hexatic membrane
has equilibrium configurations characterized by intrinsic ripples.
Ripples are well known and studied in the physics of graphene \cite{fasolino}.
While the results of Sect.\ \ref{t-f-section} predict a definite spectral dimension
when interfacing a tethered and a fluid membrane, it is clear from \eqref{sdim}
that the characterization of the fractal properties of the hexatic model require
the experimental determination of the hexatic rigidity, which enters as an effective parameter in \eqref{hexatic}.


\section{Summary.}

In a unified formalism, we obtained all the models of the KTNHY description of continuous membranes
as field theories that nonlinearly realize the Euclidean symmetries of the embedding space.
Each specific coset was constructed using the very efficient formalism of the Maurer-Cartan form
and is motivated by the physical properties that the fundamental constituents of the membrane are
assumed to posses according to the temperature of the thermal bath.
The inclusion of the physical properties of the constituents and of their crystalline structure
is essential for reproducing all the distinct phases of the KTNHY description, therefore distinguishing
our effective models of membranes from those in which the membrane is considered a fundamental object
such as string theory \cite{West:2000hr}.

Immediately identified in the effective description obtained through the Maurer-Cartan form
are both the Goldstone fields of the broken symmetries of embedding
and the relevant order parameters for the phases of each model, which in these models
have immediate physical interpretations \cite{Delacretaz:2014oxa}.
In all models, the order parameters obtained via the coset construction
coincide with those that were already well-established in the literature \cite{Bowick:1999rk},
but within the coset construction they acquire a formal definition in the sense of \cite{Weinberg:1996kr}.

The construction of the KTNHY models shows clearly the degrees of freedom
that have to be effectively integrated in the low- and intermediate-temperature regimes
to obtain effective fluid descriptions for all of them.
The effective integration of both the low-temperature crystalline phase and of the intermediate
hexatic phase give effective fluid models that possess long range interactions among
intrinsic curvatures on the membrane, which are believed to induce crinkled phases at equilibrium \cite{PL}.
The crinkled phase is determined by the presence and interactions of intrinsic ripples
of the surface characterizing the membrane \cite{fasolino}.
While the importance of role of long range interactions among displacements of the membrane
is already well known for the characterization of the flat phase of a crystalline membrane \cite{Essafi:2014dla},
we showed how these effectively manifest in a fluid description and gave a clear geometrical interpretation.

The long range interactions are induced by the thermal fluctuations,
by the melting of the crystalline structure and by the loss of local order among the fundamental constituents.
The sole presence of the long range interactions casts doubts on whether a crystalline and a fluid
descriptions may share the same phase-diagram. For a physical two-dimensional membrane
embedded in a three-dimensional space there is no conclusive proof whether the phase-diagram is shared or not.
This work can be considered as a step forward in the direction of a better understanding of this question
and a further indication that it has to be addressed non-perturbatively \cite{Codello:2011yf,Essafi:2014dla}.

Our results are especially relevant for the melting of both regular and unstructured two-dimensional crystals
into a fluid phase. In such a scenario, the crystal undergoes a crinkled phase that is characterized by an
anomalous fractal dimension of the membrane due to the presence of effective long range interactions
among the monomers.
In our computations, the spectral dimension is a genuine prediction of the formalism when the crystal does not exhibit local order
even in the crystalline phase, while it depends on the hexatic rigidity when local order is present.

\vspace{-0.3cm}

\section*{Acknowledgments.}
The author thanks M.\ Demmel, G.\ D'Odorico, S.\ Endlich and R.\ Penco for illuminating discussions
and valuable comments on the draft of this paper.
The work benefited by the hospitality of the Perimeter Institute for
Theoretical Physics during the workshop ``Low Energy Challenges for High Energy Physicists'',
where a first draft of ideas was developed.
This work is supported by the DFG within the Emmy-Noether program (Grant SA/1975 1-1).

\vfill

\end{document}